\documentstyle[aas2pp4]{article}

\newcommand{\beq}{\begin{equation}}
\newcommand{\eeq}{\end{equation}}
\def\beqa{\begin{eqnarray}}
\def\eeqa{\end{eqnarray}}
\newcommand{\lsim}{\lesssim}

\begin{document}

\title{A GALAXY X-RAY FUNDAMENTAL PLANE?}  

\author{M. Fukugita$^{1,2}$ and P. J. E. Peebles$^{1,3}$}
\affil{$^1$Institute for Advanced Study, Princeton, NJ 08540, USA}
\affil{$^2$ Institute for Cosmic Ray Research, University of Tokyo, 
Tanashi, Tokyo 188, Japan}
\affil{$^3$Joseph Henry Laboratories, Princeton University,
Princeton, NJ 08544, USA}

\begin{abstract}

We suggest the radii
and luminosities of the X-ray emitting halos of elliptical 
galaxies define a fundamental plane 
with the star velocity dispersions, as
for the corresponding optical observables. 
Since the X-ray emitting material usually is at larger 
radius than the stars this can
be interpreted as additional evidence for a 
relation between the space distributions of stellar
and dark mass, in analogy
to the Tully-Fisher relation and flat rotation 
curves for spiral galaxies. The 
picture is complicated, however, by 
ellipticals with relatively low X-ray luminosities 
that appear to have modest dark halos, 
quite unlike the standard picture for spirals. 

\end{abstract}

\section{The Fundamental Plane}

The presence of dark matter in spiral galaxies is well 
established (under conventional physics) and
correlated with the stellar mass, as in the Tully \& Fisher (1977) 
relation between the luminosity in the 
inner region, which is dominated by stars, and the 
circular velocity in the outer part that is
thought to be dominated by nonbaryonic dark matter. 
Dark matter in elliptical galaxies is less well 
explored, because there are fewer accessible probes
outside the concentration of starlight,
but observations of X-rays, globular clusters and planetary 
nebulae show an extended dark halo in 
M87 (Fabricant \&\ Gorenstein 1983) and several other 
X-ray emitting galaxies (Forman, Jones \ Tucker 1985; Bridges 1998) 

It would be interesting to know whether
elliptical galaxies have an analog 
of the Tully-Fisher relation 
between the baryonic and dark matter. The 
fundamental plane (Djorgovski \& Davis 1987; Dressler et al. 1987) 
and the Faber-Jackson (1976) relation 
apply to properties of spheroids that are likely to be 
dominated by stars; 
they seem to have little to say about the 
dark matter. The X-ray emitting material, that tends 
to be at larger radii than the stars, has luminosity
that correlates with the starlight, 
roughly as $L_x\sim L_o^{1.8}$, 
but with an order of magnitude scatter 
in $L_x$ at given $L_o$ (Fabbiano, Gioia \& Trinchieri 1989;
Donnelly, Faber \& O'Connell 1990;
Eskridge, Fabbiano \&\ Kim 1995).
Mathews \& Brighenti (1998)
propose a tighter relation, 
$L_x/L_o \propto (r_{x}/r_o)^{0.6\pm 0.3}$, where $r_o$
and $r_x$ are the optical and X-ray half-luminosity radii. 
In this paper we present evidence for an even 
tighter fundamental plane relation in X-ray properties, 
 with a looser connection to optical observables,
and we consider possible implications for dark halos. 

We write the fundamental plane relation as 
\beq
L\propto\sigma ^a r^b,
\label{eq:fp}
\eeq
where $r$ is the half-luminosity radius for the optical or X-ray 
luminosity $L$, and $\sigma$ is the 
central line of sight star velocity dispersion. 
For our purpose this is simpler than 
the $\kappa$-space convention of Burstein et al. (1997). 
If the mass-to-light ratio were constant
in an isothermal sphere with small core radius 
the parameters 
would be $a=2$ and $b=1$. 
For the proposed X-ray relation 
it would be better to use the X-ray plasma temperature, but 
that might best await further  
measurements, perhaps with the CHANDRA satellite.

Scatter plots around equation~(\ref{eq:fp}) are shown in 
Figure 1 for
$a=2$ and $b=1$, and in Figure~2 for $a$ and $b$ 
adjusted to
minimize the scatter, as discussed below. 
We  choose units for the 
data set in each panel so $\log L$, 
$\log\sigma$, and $\log r$ have zero medians.
The lines are equation (1). 
The filled squares are the 11 galaxies 
for which Mathews \&\ Brighenti (1998) find there are 
suitable X-ray images from which they can estimate
half-luminosity radii. The optical data 
for this sample are from Bender, Burstein \&\ Faber (1993). 
The open symbols
are the 19 ellipticals with 
measured X-ray luminosities as well as optical quantities, 
 from the compilations of Eskridge, Fabbiano \&\ Kim (1995) 
and Bender, Burstein \&\ Faber (1992). The data file we use is 
available on request. 

An optical fundamental plane relation is shown in the 
upper left panels of the figures. The relation for the 
X-ray luminosity $L_x$ and half-luminosity radius $r_x$, 
with the optical velocity 
dispersion, is in the lower right-hand panel. 
The other panels combine the optical half-light radius 
$r_o$ with $L_x$, and $L_o$ with $r_x$. 

A least squares measure of the scatter may be unduely 
affected by non-Gaussian fluctuations, so we use medians. With
the normalization to zero median values,
a measure of the scatter is the median of 
\beq
A(a,b) = |\log L(i) - a\log\sigma (i) - b\log r(i)|,
\label{eq:A}
\eeq
in the sample of $i_x=11$ galaxies with X-ray radii 
or $i_x = 30$ galaxies with X-ray luminosities but 
not necessarily measured X-ray radii. 
If $L$, $\sigma$ and $r$ were statistically independent
the expected frequency distribution of the $A(a,b)$ 
would be the same as the distribution of 
\beq
B(a,b) = |\log L(j) - a\log\sigma(k) - b\log r(n)|,
\label{eq:B}
\eeq
where $j$, $k$, and $n$ are drawn 
independently at random from the integers $1\leq i\leq i_x$.
The ratio 
\beq
R(a,b) = {\rm med}\, A(a,b)/{\rm med}\, B(a,b)
\label{eq:R}
\eeq
is a measure of the correlation. We use the median of 5000 
samples of $B$, and find the 
$a$ and $b$ that minimize $R(a,b)$ by numerical search.
Table~1 lists the results.

For $L_o$, $r_x$ there are several minima, none sharply 
defined; Table 1 and Figure 2 show a representative case. 
For $L_x$, $r_o$ there is another local minimum at
$a = 0.1$, $b=1.9$ with $R = 0.43$, almost as small as in 
the table, but the fit does not look as good as Figure 2.
When  medians are replaced with means the minima are 
broader but at similar 
positions. With mean square values the 
minima are still broader and $a$ and $b$ at the 
minima are substantially different for all 
but $L_o$, $r_o$. 

The fifth column in Table 1 is $R$ at $a=2$ and $b=1$, for 
constant $M/L$. Under the postulate of statistical 
independence of $L$, $r$ and $\sigma$, the 
probability that $R$ is less than 
the data value $R(2,1)$ is the probability 
that more than half of $i_x$ random selections of $B$
are less than the median value of $A(2,1)$.  
This probability, in the last column of the 
table,\footnote{A computation of
probabilities or of statistical uncertainties 
of $a$ and $b$ when these parameters are adjusted 
would be desirable but impractical
by the present method because each random pseudo-data
set would require a separate parameter search.}
is computed using the binomial distribution 
and the fraction of the $i_x^3$ values of 
$B$ that are less than ${\rm med}\, A(2,1)$. The
probabilities seem consistent with the visual impression
 of Figure~1: one sees convincing correlation for
$L_o$, $r_o$, little evidence of it
for $L_o$, $r_x$, and the impression that the 
$L_x$, $r_o$ case could be improved by
adjusting $a$ and $b$, as in Figure~2. 
The correlation in the $L_x$, $r_x$ case looks promising 
albeit not compelling, at about
the 3\%\ level. The very discrepant galaxy in this panel, 
NGC5044, also is the largest anomaly in the 
radius-luminosity correlation of Mathews \&\ Brighenti (1998). 
This galaxy fits the optical fundamental plane; the problem
is not that the distance somehow is wrong. 

Adjustment of the parameters 
to minimize $R$ for $L_o$, $r_o$ 
only slightly reduces 
the scatter. The scatter is also similar 
for the optical fundamental 
plane parameters $a=1.51$, $b=0.78$ 
 from J$\phi$rgensen, Franx \& Kj\ae rgaard (1996). 
That is, our data set for $L_o$, $r_o$ shows 
a reasonably tight fundamental plane relation for a substantial 
range of values of the power law index of $\sigma$, 
$1\lsim a\lsim 2$, when $b$ is adjusted to a value near unity. 
The parameters for $L_x$, $r_x$ are not tightly constrained
either, as one sees from the similar quality of fit 
in the two figures. 
Our sample is biased 
by the restriction to galaxies with detected X-ray 
luminosity, and it may have some
distance errors, but the selection does yield a 
tight relation among the optical observables
and a promising relation among X-ray observables.

\section{Discussion}

A closer test of the proposed X-ray fundamental plane
awaits a larger 
sample, idealy with X-ray temperatures, but the 
present case seems good enough
to motivate preliminary discussion. 

The $L_o$, $r_o$ and $L_x$, $r_x$ relations in the 
constant $M/L$ model 
imply $L_x/L_o\propto r_x/r_o$. This agrees with the Mathews \&\
Brighenti (1998) relation at one standard deviation. 
We are proposing a relation among the velocity dispersion  
and the X-ray luminosity and radius that 
is tighter than 
similar combinations of $\sigma$ with $L_x$, $r_o$ 
or $L_o$, $r_x$.

The optical fundamental plane in our sample
is roughly consistent with the picture that the 
spheroid mass is dominated by old stars 
with close to universal mass-to-light ratio (as in
examples of $M/L$ in ellipticals with cold 
gaseous disks; Pizzella et al. 1997). 
Among all high surface density early-type galaxies 
$M/L$  correlates with $M$ 
(Burstein et al. 1997 and references therein), but the 
constant $M/L$ model could be an adequate picture in the 
mass range of X-ray luminous ellipticals. This model 
requires close to universal star age and mass 
functions, to account for near constant $M/L$ dominated
by stars, but it says nothing about homology of the 
spheroids.

 From the X-ray data we can argue for the following 
constraint on homology. The X-ray emitting material,
whether accreted from outside a galaxy or in part shed 
by stars, settles at characteristic radius $r_x$. 
The halo circular velocity $v_x$ at $r_x$ determines the
temperature $T_x$. That with the X-ray emitting mass $M_x$ 
determines the X-ray luminosity $L_x$. We are suggesting $r_x$ 
and $L_x$ are tightly related to the star velocity 
dispersion $\sigma$. Since $r_x$ is not tightly 
correlated with the star half-light radius 
$r_o$ this requires that $v_x$ is 
well correlated with $\sigma$, 
meaning that stellar and dark mass distributions
are correlated. This is an 
analog to the 
Tully-Fisher relation between stellar and dark mass in spirals. 
Our argument also implies that
$r_x$ is tightly correlated with $M_x$, though both are
only loosely related to the optical observables. 
This is curious enough to be treated 
with caution, but the Tully-Fisher
relation --- the ``conspiracy'' relating baryonic
and dark matter in spirals --- also is curious.
The dark matter ``conspiracy'' in ellipticals is
more directly probed by comparing  star velocity
dispersions and X-ray temperatures. 
Davis \&\ White (1996) find evidence for this correlation,
and Brighenti \&\ Mathews (1997) for a universal form for the 
X-ray temperature run. Further exploration with CHANDRA
will be of considerable interest. 

If the proposed analog to Tully-Fisher is established it 
is not likely to be the whole story,
because some elliptical galaxies 
seem to be endowed with relatively little dark matter. 
For M105
Ciardullo, Jacoby \&\ Dejonghe (1993) find $M/L_B\sim 7$ 
solar units at 10 kpc radius, from the redshift 
distribution of planetary nebulae. This is
consistent with stellar values for an old population
without dark matter. Schneider's (1991) analysis of the 
HI ring that encircles this galaxy and NGC3384
yields $M/L_B\sim 25$ at semi-major axis 100 kpc. 
Unless this is a near face-on S0 galaxy, and a near face-on 
ring accidently matches Schneider's excellent  fit of
angular position and radial velocity to a Keplerian 
orbit, this is a marked contrast with what we know 
about spiral galaxies. Several other examples are in 
the review by Bridges (1998). In particular, in NGC5128 
(the host of Cen A) Peng et al. (as quoted  
by Bridges 1998) find that if the anisotropy
of the planetary velocities is independent 
of radius the projected mass 
increases with radius only 
to 20 kpc, with $M/L_B$ as low as 15 at 50 kpc radius,
again not much more than is in the stars. 
These ellipticals with low $M/L$ are not associated with
strong X ray emitting haloes.   
Well-established examples of elliptical galaxies with prominent
dark mass halos, as reviewed by Bridges (1998), 
tend to be strong X-ray sources. 

We have proposed that the dark mass halos of X-ray luminous 
ellipticals fit an  
X-ray fundamental plane relation, and that this implies
an analog of the Tully-Fisher relation between baryonic
and dark matter observables. The evidence is that another class
of ellipticals have normal stellar spheroids but low X-ray
luminosities and inconspicuous dark matter halos. If confirmed, 
these are interesting clues and challenges to the theory
of galaxy formation. 

\acknowledgements

We are grateful to David Hogg and Penny Sackett for 
discussions, and to David Burstein and Bill Mathews for their 
careful reading and helpful comments on our paper. 
This work was supported in part at the Institute for Advanced Study 
by the Alfred P. Sloan Foundation and the Raymond and Beverly
Sackler Fellowship.

\bigskip
\def\hrulefil{\leaders\hrule height0.6pt\hfill\quad}
\def\hrulefill{\leaders\hrule height0.6pt\hfill}
\centerline{TABLE 1}
\centerline{PARAMETER FITS}
\hbox to \hsize{\hss\vbox{
\vskip 6pt 
\hrule height 0.6pt
\vskip 2pt
\hrule height 0.6pt
\halign{ \hfil #\hfil & \ \hfil #\hfil\ & \ \hfil #\hfil\
 &\hfil #\hfil & \hfil #\hfil & \hfil #\hfil \cr
\noalign{\vskip 4pt}
Variables & a & b & R(a,b) & R(2,1) & $P(2,1)$\cr
\noalign{\vskip 4pt}
\noalign{\hrule height 0.6pt}
\noalign{\vskip 4pt}
  $L_x$, $r_x$  &  0.8 & 0.6 & 0.14 & 0.46 & 0.03 \cr
  $L_o$, $r_x$  &  2.2 & 0.2 & 0.38 & 1.15 & 0.7 \cr 
  $L_x$, $r_o$  &  3.9 & 2.3 & 0.41 & 0.79 & 0.14 \cr 
  $L_o$, $r_o$  &  1.3 & 1.0 & 0.18 & 0.30 & 0.000018 \cr 
\noalign{\vskip 4pt}
\noalign{\hrule height 0.6pt}
}}\hss}

\bigskip

\plotone{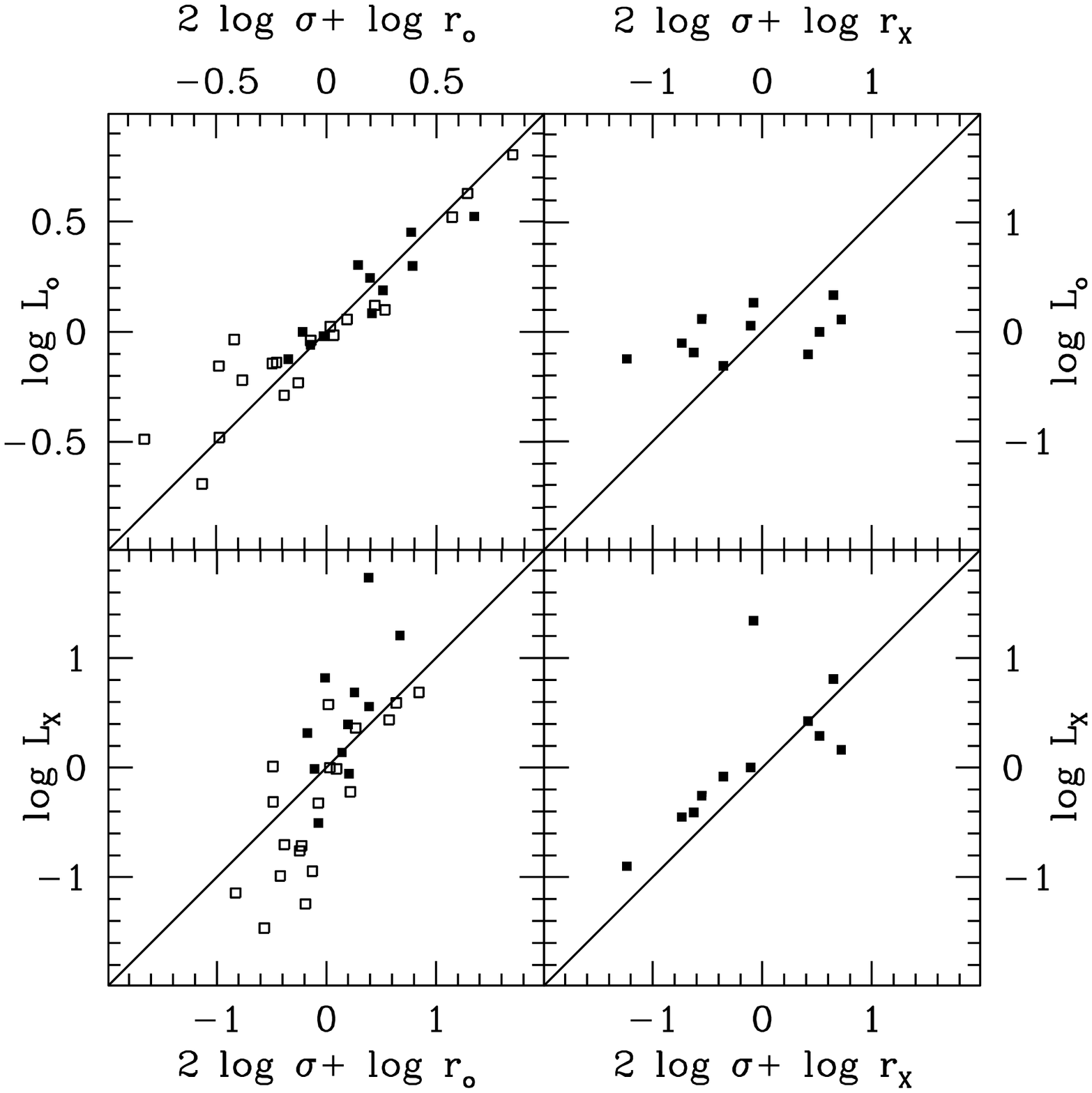}
\figcaption[figure1ii.ps]{Fundamental plane relations
with $a=2$ and $b=1$ in equation~(\ref{eq:fp}). Units in 
each panel are
chosen so $\log L$, $\log r$ and $\log\sigma$ have
zero median values.}

\bigskip

\plotone{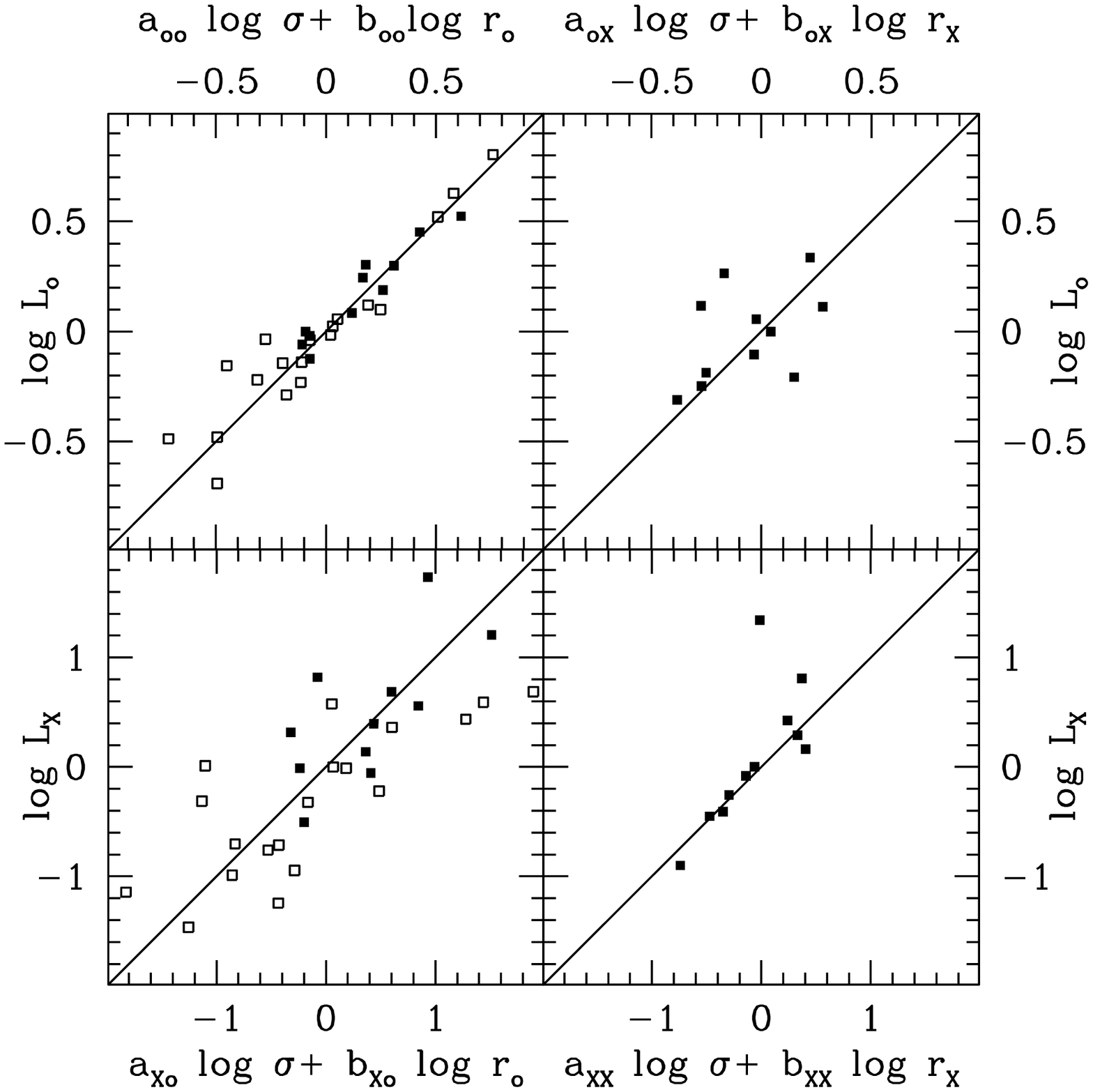}
\figcaption[figure2ii.ps]{Fundamental plane relations 
with $a$ and $b$ adjusted to the values in Table~1
that minimize the scatter as measured by 
equation~(\ref{eq:R}).} 

\end{document}